\documentclass[aps,amsmath,amssymb,nofootinbib]{revtex4} 
\usepackage{graphicx} 
\usepackage{amsmath}
\usepackage{amsfonts,amsbsy}
\usepackage{amssymb}

\def\be{\begin{equation}}
\def\ee{\end{equation}}
\def\bea{\begin{eqnarray}}
\def\eea{\end{eqnarray}}

\begin{document}

\title{$\cos(4\phi)$ azimuthal anisotropy in small-$x$ DIS dijet
  production beyond the leading power TMD limit}

\author{Adrian Dumitru}
\affiliation{Department of Natural Sciences, Baruch College, CUNY,
17 Lexington Avenue, New York, NY 10010, USA}
\affiliation{The Graduate School and University Center, The City
  University of New York, 365 Fifth Avenue, New York, NY 10016, USA}

\author{Vladimir Skokov}
\affiliation{RIKEN/BNL Research Center, Brookhaven National
  Laboratory, Upton, NY 11973, USA}

\begin{abstract}
We determine the first correction to the quadrupole operator in
high-energy QCD beyond the TMD limit of Weizs\"acker-Williams and
linearly polarized gluon distributions. These functions give rise to
isotropic resp.\ $\sim\cos 2\phi$ angular distributions in DIS dijet
production. On the other hand, the correction produces a $\sim \cos
4\phi$ angular dependence which is suppressed by one additional power
of the dijet transverse momentum scale (squared) $P^2$.
\end{abstract}

\maketitle

\section{Introduction}

We consider (inclusive) production of a $q\bar{q}$ dijet at leading
order in high-energy (small-$x$) Deep Inelastic Scattering (DIS) of an
electron off a proton or nucleus. The average transverse momentum of
the jets is denoted as $P$ and the transverse momentum imbalance is
$q$. In the ``correlation limit'' of roughly back to back
jets~\cite{Dominguez:2011wm} one has $P^2\gg q^2$. In this limit the
leading contribution (in terms of powers of $1/P^2$) to the cross
section can be obtained from Transverse Momentum Dependent (TMD)
factorization. For a recent review of TMD factorization see
Ref.~\cite{Angeles-Martinez:2015sea}. It predicts a distribution for
linearly polarized gluons in an unpolarized
target~\cite{Mulders:2000sh,Meissner:2007rx} which gives rise to
$\sim\cos 2\phi$ asymmetries in dijet
production~\cite{Boer:2009nc,Metz:2011wb,DQXY} and in other
processes~\cite{Boer:2010zf,Qiu:2011ai}. The azimuthal angle $\phi$ is
the angle between the transverse momentum vectors $P$ and
$q$\footnote{To avoid cluttering of notation we do not write vector
  arrows on 2d vectors. In this paper essentially all transverse
  coordinates and momenta are 2d vectors and their magnitudes are
  written as $|P|$ or $\sqrt{P^2}$ etc.}. At small $x$ the
distribution of linearly polarized gluons $xh^{(1)}(x,q^2)$ is
expected to be comparable in magnitude to the conventional
Weizs\"acker-Williams gluon distribution $xG^{(1)}(x,q^2)$ for $q^2$
of order the saturation momentum scale $Q_s^2$ of the
target~\cite{Dumitru:2015gaa}. These could be measured at a future
electron-ion collider (EIC)~\cite{Boer:2011fh}. However, the
experimental collaborations have requested an
estimate~\cite{exp_cos4phi} of $\langle\cos 4\phi\rangle$ since this
may constitute a background for $\langle\cos 2\phi\rangle$ which is
generated by $xh^{(1)}(x,q^2)$.

Corrections to TMD factorization appear beyond the leading order in
$1/P^2$. In what follows we derive the operator form of the correction
in Eqs.~(\ref{Eq:G22},\ref{Eq:G31a},\ref{Eq:G31b}), and explicit
expressions for the expectation values in a large-$N_c$ Gaussian
theory in Eq.~(\ref{eq:QG4thO}), and we show that it
leads to a new $\sim\cos 4\phi$ azimuthal harmonic.

The remainder of the paper is organized as follows. In section
\ref{sec:Qcorr_4phi} we derive the operator corresponding to the
correction to the (leading power) TMD approximation of the quadrupole. In
sec.~\ref{sec:Gauss} we use a Gaussian (and large-$N_c$) approximation
to obtain explicit expressions of this correction in terms of the
two-point function of the Gaussian theory. The correction to the dijet
cross section is worked out in sec.~\ref{sec:DijetXsec} where we also
provide a first qualitative estimate of the magnitude of $\langle \cos
4\phi\rangle$ relative to $\langle \cos 2\phi\rangle$. We close with a
brief summary in sec.~\ref{sec:Summary}.

\section{Extracting the azimuthal angular components of the
  quadrupole operator} \label{sec:Qcorr_4phi}

The production of a quark anti-quark dijet at small $x$ in DIS
involves the following expectation value of Wilson
lines~\cite{Dominguez:2011wm}:
\be
{\cal Q}_x(x_1,x_2;x_2^{\prime },x_1^{\prime})
 = 1+S^{(4)}_x(x_1,x_2;x_2^{\prime },x_1^{\prime})
-S^{(2)}_x(x_1,x_2)-S^{(2)}_x(x_2^{\prime },x_1^{\prime })  ,
\ee
where
\be
S^{(2)}_x(x_1,x_2) = S^{(2)}_x((x_1-x_2)^2)
\equiv \frac{1}{N_c} \left< {\rm Tr}\,
V^\dagger(x_2)  V(x_1) \right>_x
\ee
is the dipole S-matrix evolved to light-cone momentum
fraction $x$; we omit this subscript on the $\langle\cdot\rangle$
field configuration averages from now on. The field of the target is
taken in covariant gauge. Also, $x_1$ and $x_2$ denote the 2d
transverse coordinates of the fundamental Wilson lines corresponding
to the quark and the anti-quark, respectively. The saturation momentum
scale where the fields of the target become non-linear is
conventionally defined implicitly through $S^{(2)}(2/Q_s^2) = 1/\sqrt{e}$.

The quadrupole operator is given by a single trace over four Wilson
lines,
\be
S^{(4)}(x_1,x_2;x_2^{\prime },x_1^{\prime})
\equiv \frac{1}{N_c} \left< {\rm Tr}\,
V^\dagger(x_2)  V(x_1) V^\dagger(x_1^{\prime})  V(x_2^{\prime})
\right> ~.
\ee
${\cal Q}_x(x_1,x_2;x_2^{\prime },x_1^{\prime})$ vanishes in the
coincidence limits $x_1\to x_2$ or $x_1^{\prime}\to x_2^{\prime}$.

In the so-called ``correlation limit''~\cite{Dominguez:2011wm,DQXY} of
roughly back to back jets it is useful to introduce
\be
{u} = {x}_1 - {x}_2; \quad {v} =   \frac12 \left(
{x}_1 + {x}_2 \right)~,
\ee
and similar for the primed coordinates. In this limit one expands
${\cal Q}$ in powers of $u$ and $u^\prime$,
\be
{\cal Q} =
u_{i} u_{j}^\prime {\cal G}^{i,j}({v}, {v}^\prime) 
+ u_{i} u_{j}^\prime u_{k}^\prime u_{l}^\prime {\cal G}^{i,jkl}({v}, {v}^\prime) 
+ u_{i} u_{j} u_{k} u_{l}^\prime {\cal G}^{ijk,l}({v}, {v}^\prime) 
+ u_{i} u_{j} u_{k}^\prime u_{l}^\prime {\cal G}^{ij,kl}({v}, {v}^\prime) 
	\label{Eq:Qexpansion} + \cdots~.
\ee
Ref.~\cite{DQXY} performed the expansion to order ${\cal
  O}(uu^\prime)$ from where one obtains the Weizs\"acker-Williams (WW)
gluon distribution. It is proportional to
\begin{equation}
  {\cal G}^{i,j}  ({v}, {v}^\prime) =
  - \frac{1}{N_c} 
  \langle {\rm Tr} \,
  V^\dagger({v}) \partial_i V({v}) 
  V^\dagger({v}^\prime) \partial_j V({v}^\prime) 
\rangle~.
	\label{Eq:Q_11}
\end{equation}
This is a two-point correlator of the target field transformed to
light-cone gauge and so defines a gluon distribution. Its Fourier transform,
\begin{equation}
  x G^{ij}_{\rm WW}(x,{q})  \equiv  \frac{2 N_c}{\alpha_s}  {\cal
    G}^{i,j}({q}) = - \frac{2}{\alpha_s}
  \int \frac{\text{d}^{2}v}{(2\pi)^{2}}\frac{\text{d}^{2}v^{\prime }}{(2\pi )^{2}}
	e^{-i {q}\cdot ({v}-{v}^{\prime})} 
	\langle  {\rm Tr}\,
	V^\dagger({v}) \partial_i V({v}) 
	V^\dagger({v}^\prime) \partial_j V({v}^\prime) 
\rangle  \label{eq:FTxGij}
\end{equation}
can be projected onto its diagonal and traceless parts 
\begin{equation} \label{eq:WW_G_h}
	x G^{ij}_{\rm WW}(x,{q})  = \frac12 \delta_{ij}\, xG^{(1)}(x,q^2) 
        - \frac12 \left(\delta_{ij} - 2 \frac{q_i q_j}{q^2}
	 \right) xh^{(1)}(x,q^2)~.
\end{equation}
The conventional WW gluon distribution $xG^{(1)}(x,q^2)$ leads to a
dijet cross section which is isotropic in $\phi$, i.e.\ in the angle
between the dijet transverse momentum imbalance $q$ and the average
transverse momentum $P$.

In Eq.~(\ref{eq:WW_G_h}) the distribution of linearly polarized gluons
is denoted as $xh^{(1)}(x,q^2)$. This function has been computed
within the McLerran-Venugopalan (MV) model of semi-classical gluon
fields~\cite{McLerran:1993ni} in Refs.~\cite{DQXY,Metz:2011wb}, and
its QCD quantum evolution to small-$x$ has been determined in
Ref.~\cite{Dumitru:2015gaa}. A non-vanishing $xh^{(1)}(x,q^2)$ gives
rise to a $\sim\cos(2\phi)$ azimuthal anisotropy of the dijet cross
section which is long range in the rapidity asymmetry of the
dijet~\cite{Dumitru:2015gaa}.

In this paper we extend the expansion to fourth order in $u$ and/or
$u^\prime$ as indicated in Eq.~(\ref{Eq:Qexpansion}). At quartic
order,
\begin{eqnarray}
{\cal G}^{ij,mn}(v,v^\prime) &=& \frac1{16N_c} \langle {\rm Tr} \, \left[
	V^\dagger({v})\partial_i \partial_j  V({v}) +   
	(\partial_i \partial_j V^\dagger({v}))  V({v}) 
	\right]
 \left[(\partial_m \partial_n   V^\dagger({v}^\prime)) V({v}^\prime) +   
	V^\dagger({v}^\prime)  \partial_m \partial_n V({v}^\prime)
        \right] \rangle \label{Eq:G22} ~, 
\\
{\cal G}^{ijm,n}(v,v^\prime) &=& - \frac1{24N_c}
	\langle {\rm Tr}\,
	\left[V^\dagger({v})\partial_i \partial_j \partial_m V({v}) 
	+ 3 (\partial_i \partial_j V^\dagger({v})) \partial_m V({v})  
	\right] 
	V^\dagger({v}^\prime) \partial_n V({v}^\prime)
	\rangle \label{Eq:G31a} ~,\\
{\cal G}^{n,ijm}(v,v^\prime) &=& - \frac1{24N_c}
	\langle {\rm Tr}\,
	 \left[V^\dagger({v}) \partial_n V({v})\right]
	\left[
	V^\dagger({v}^\prime)\partial_i \partial_j \partial_m  V({v}^\prime) 
	+ 3 (\partial_i \partial_j V^\dagger({v}^\prime))  \partial_m V({v}^\prime) 
	\right]
	\rangle~.   \label{Eq:G31b}
\end{eqnarray}
These expressions have been simplified by taking advantage of the
symmetries in Eq.~(\ref{Eq:Qexpansion}). Their Fourier transforms are
performed as for the WW distribution in Eq.~(\ref{eq:FTxGij}) above
and the resulting tensors can be decomposed as follows:
\begin{eqnarray}
\frac{2N_c}{\alpha_s}  {\cal G}^{ijmn}(x,{q}^2) = 
\mathfrak{P}_1^{ijkm} \Phi_0(x,{q}^2) 
+ \mathfrak{P}_2^{ijkm} \Phi_1(x,{q}^2) 
-  \mathfrak{P}_3^{ijkm} \Phi_2(x,{q}^2) ~,
	\label{Eq:Decomposition}
\end{eqnarray}
where
\begin{eqnarray}
{\cal G}^{ijmn}(x,{q}^2) &=& {\cal G}^{i,jmn}(x,{q}^2) + {\cal G}^{ijm,n}(x,{q}^2) -
\frac{2}{3}{\cal G}^{ij, mn}(x,{q}^2) ~, \label{eq:Gijmn}\\
\mathfrak{P}_1^{ijmn} &=& \frac{1}{2\sqrt{6}} \left(  \delta_{ij}
\delta_{mn} +  \delta_{im} \delta_{jn} +  \delta_{jm} \delta_{in}
\right)~, \\
\mathfrak{P}_2^{ijmn} &=& - \frac{1}{6\sqrt{2}} \left(  
\delta_{ij} \Pi_{mn}  + 
\delta_{mn} \Pi_{ij}  + 
\delta_{im} \Pi_{jn} +
\delta_{jn} \Pi_{im} +
\delta_{jm} \Pi_{in} + 
\delta_{in} \Pi_{jm}  
\right)  ~,\\ 
\mathfrak{P}_3^{ijmn} &=& - \frac{1}{6\sqrt{2}} \left(  \delta_{ij}
\delta_{mn} +  \delta_{im} \delta_{jn} +  \delta_{jm} \delta_{in}
- 2 (
\Pi_{ij} \Pi_{mn} +  \Pi_{im} \Pi_{jn} +  \Pi_{jm} \Pi_{in}
)
\right)  ~,\label{Eq:P} \\
\Pi_{ij} &=& \delta_{ij} - \frac{2 q_i q_j}{q^2} ~.\label{Eq:Pi}
\end{eqnarray}
The function ${\cal G}^{ijmn}(x,{q}^2)$ as introduced in
Eq.~(\ref{eq:Gijmn}) appears in the dijet cross section, see
section~\ref{sec:DijetXsec} below.

The projectors are normalized so that $\mathfrak{P}_i^2 =
1$ for $i=1,2,3$ and they satisfy
\begin{eqnarray}
P_i P_j P_m P_n \mathfrak{P}_1^{ijmn} &=&  \frac12 \sqrt{\frac{3}{2}} P^4~,\\
P_i P_j P_m P_n \mathfrak{P}_2^{ijmn} &=& \frac1{\sqrt{2}} P^4 \cos 2\phi ~,\\
P_i P_j P_m P_n \mathfrak{P}_3^{ijmn} &=&  \frac{1}{2\sqrt2} P^4 \cos 4\phi~. 
\end{eqnarray}
Hence, the parity of $\mathfrak{P}_i$ under $\phi\to\phi+\pi/2$ is $(-)^{i-1}$.

In what follows we shall focus on $\Phi_2(x,{q}^2)$ which determines
the amplitude of the $\sim\cos\, 4\phi$ contribution to dijet
production,
\be \label{eq:Phi2_P_Gijmn}
\Phi_2(x,{q}^2) = - \frac{2 N_c}{\alpha_s}  \mathfrak{P}_3^{ijmn}
    {\cal G}^{ijmn}(x,q^2)~.
\ee
The first two terms from Eq.~(\ref{Eq:Decomposition}) only
contribute corrections (suppressed by $\sim 1/P^2$) to the isotropic and
``elliptic'' ($\sim\cos\, 2\phi$) contributions.

Equation (\ref{eq:Phi2_P_Gijmn}) is the final result of this section. It
expresses the correlation function $\Phi_2(x,{q}^2)$ which determines
the $\sim\cos\, 4\phi$ asymmetry in terms of a combination of
correlation functions of Wilson lines written in
Eqs.~(\ref{Eq:G22},\ref{Eq:G31a},\ref{Eq:G31b}).

%
\section{Gaussian approximation} \label{sec:Gauss}

In this section we compute the correlator $\Phi_2(x,{q}^2)$
analytically in the Gaussian and large-$N_c$ approximations. The
Gaussian theory is believed to be a good approximation at small
$x$~\cite{smallx_Gauss} unless, perhaps, the contribution from
so-called ``pomeron loops'' is large~\cite{KL_pomloops}.  This has
been confirmed explicitly by a numerical
analysis~\cite{Dumitru:2011vk}. Note, however, that
Ref.~\cite{Dumitru:2011vk} did not test configurations corresponding
to large $v-v^\prime$ and small $u$, $u^\prime$ as required for the
present analysis.

At a Gaussian fixed point the theory is defined in terms of the
two-point function
\bea
g^2 \langle A^{-a}(z_1^+,z_1)\, A^{-b}(z_2^+,z_2)\rangle &=&
\delta^{ab}\, \delta(z_1^+-z_2^+)\, \mu^2(z^+)\, L_{z_1z_2} ~, \\
L_{z_1z_2} &=& g^4 \int d^2z\, G_0(z_1-z)G_0(z_2-z)~,\\
G_0(z) &=& \int \frac{d^2k}{(2\pi)^2} \frac1{k^2+\Lambda^2_{\rm IR}}
e^{ik\cdot z} = \frac{1}{4\pi} \ln \frac{1}{z^2 \Lambda^2_{\rm IR}}~.
\eea
$\Lambda_{\rm IR}$ regularizes the long-distance 2d Coulomb singularity and we
restrict to $z^2 \Lambda^2_{\rm IR}\ll1$.
This leads to the dipole S-matrix
\be
S^{(2)}(x_1,x_2) = \exp\left( -\frac1{2} C_F \Gamma((x_1-x_2)^2)\right) ~,
\ee
where
\be
\Gamma(r^2) = 2 \left(L(0) - L(r^2)\right)~.
\ee

In the large-$N_c$ limit ${\cal Q}$ as defined in
\eqref{Eq:Qexpansion} can be written in the
Gaussian theory as~\cite{Blaizot:2004wv,Dominguez:2011wm}
\begin{eqnarray}
{\cal Q}^{G} &=& 1 + e^{-\frac{C_F}{2} \left[ \Gamma({x}_1 -
    {x}_2)   +  \Gamma({x}^\prime_2 - {x}^\prime_1)
    \right]}
-  e^{-\frac{C_F}{2} \left[ \Gamma({x}_1 - {x}_2)   \right]}
-  e^{-\frac{C_F}{2} \left[ \Gamma({x}^\prime_2 - {x}^\prime_1)   \right]}\\
&&- 
\frac{ 
\Gamma({x}_1 - {x}^\prime_1) -
\Gamma({x}_1 - {x}^\prime_2) +
\Gamma({x}_2 - {x}^\prime_2) - 
\Gamma({x}_2 - {x}^\prime_1) 
}
{
\Gamma({x}_1 - {x}^\prime_1) -
\Gamma({x}_1 - {x}_2) +
\Gamma({x}_2 - {x}^\prime_2) - 
\Gamma({x}^\prime_2 - {x}^\prime_1) 
}
\left( 
e^{ -\frac{C_F}{2} \left[ \Gamma({x}_1 - {x}_2)   +  \Gamma({x}^\prime_2 - {x}^\prime_1)   \right]} - 
e^{ -\frac{C_F}{2} \left[ \Gamma({x}_1 - {x}^\prime_1)   +  \Gamma({x}^\prime_2 - {x}_2)   \right]} 
\right)~ \notag .
\label{Eq:Q_MV}
\end{eqnarray}
We now express $x_1$, $x_2$, $x_1^\prime$, $x_2^\prime$ in terms of $u$,
$u^\prime$, $v$, $v^\prime$ and expand in powers of $u$ and
$u^\prime$. The leading contribution at quadratic order is
\begin{equation}
  {\cal G}^{i,j}(r^2) = 
\left(1-[S^{(2)}(r^2)]^2\right)
\left( \delta_{ij} \frac{\Gamma^{(1)}(r^2) }{\Gamma(r^2)} 
	+ 2 r_i r_j  \frac{\Gamma^{(2)}(r^2) }{\Gamma(r^2)}  
	\right) ~, 	\label{Eq:Q11}
\end{equation}
where 
\begin{equation}
  \Gamma^{(n)}(r^2)  = \frac{d^n \Gamma(r^2)}{d (r^2)^n} ~~~,~~~
r\equiv v-v^\prime~.
\end{equation}
From this one obtains the gluon distributions 
\begin{equation}
  xh^{(1)}(x,q^2) = \frac{2N_c }{\alpha_s} \Pi_{ij} {\cal G}^{i,j}(x,q^2)
  =  \frac{4 N_c}{\alpha_s} \frac{S_\perp}{(2\pi)^3} 
	\int d|r|\, |r|^3  J_2( |q|\,  |r| )   \left(1-[S^{(2)}(r^2)]^2\right)
        \frac{\Gamma^{(2)} (r^2)}{\Gamma (r^2)}  
\end{equation}
and 
\begin{equation}
  xG^{(1)}(x,q^2) = \frac{2N_c }{\alpha_s} \delta_{ij} {\cal
    G}^{i,j}(x,q^2)
  =  \frac{4 N_c}{\alpha_s} 
	\frac{S_\perp}{(2\pi)^3} 
	\int d|r|\, |r|  J_0( |q|\, |r| )   \left(1-[S^{(2)}(r^2)]^2\right)
        \left(   \frac{\Gamma^{(1)} (r^2)}{\Gamma (r^2)} +  r^2
        \frac{\Gamma^{(2)} (r^2)}{\Gamma (r^2)} \right)~.
\end{equation}
$S_\perp$ denotes a transverse area.

In the MV model, in leading $\log 1/r^2 \Lambda^2_{\rm IR}\gg1$ approximation,
\begin{equation}
  \Gamma(r^2) = \frac{Q_s^2}{4C_F} r^2\,  \log\frac{1}{r^2 \Lambda^2_{\rm IR}}~,
\end{equation}
where $Q_s$ denotes the saturation momentum. Note that the logarithmic
factor in $\Gamma(r^2)$ ensures that the Fourier transform of the
dipole S-matrix is a power-law at high momentum, rather than a
Gaussian; it also leads to a non-vanishing second derivative of
$\Gamma(r^2)$ w.r.t.\ $r^2$ to generate the distribution of linearly
polarized gluons, $xh^{(1)}(x,q^2)$:
\bea
xh^{(1)}(x,q^2) &=& \frac{N_c S_\perp}{2 \pi^3 \alpha_s} 
	\int d|r|\, 
        |r| J_2(|r|\, |q| ) \left[ 1-\exp\left(  -\frac{Q_s^2 r^2}{4}
          \log \frac{1}{r^2 \Lambda_{\rm IR}^2}  \right) \right]
        \frac{1}{r^2 \log \frac{1}{r^2\Lambda^2_{\rm IR}}} ~, 
  \label{eq:Gauss_MV_xh1}\\
xG^{(1)}(x,q^2) &=& \frac{N_c S_\perp}{2 \pi^3 \alpha_s} 
	\int d|r|\, 
        |r| J_0(|r|\, |q| ) \left[ 1-\exp\left(  -\frac{Q_s^2 r^2}{4}
          \log \frac{1}{r^2 \Lambda_{\rm IR}^2}  \right) \right]
        \frac{1}{r^2 }
   \label{eq:Gauss_MV_xG1}
\eea
which has been obtained previously in Refs.~\cite{DQXY,Metz:2011wb}.

At fourth order in $u$ and/or $u^\prime$ we find the following
additional contribution to ${\cal Q}^G$:
\bea
& & 
\frac14 C_F^2 \Gamma(u^2)\Gamma(u^{\prime 2}) + 
\left[1 -\left[S^{(2)}(r^2)\right]^2\right]
 \left[
   u\cdot u^\prime \left[ \frac14(u^2+u^{\prime
       2})\frac{\Gamma^{(2)}(r^2)}{\Gamma(r^2)} 
     + \frac12\left((u\cdot r)^2 + (u^\prime\cdot
     r)^2\right)\frac{\Gamma^{(3)}(r^2)}{\Gamma(r^2)} 
     \right] \right.\nonumber\\
& & \left. 
\hspace{5cm} + u\cdot r\; u^\prime\cdot r\, \left[
  \frac12\left(u^2+u^{\prime
    2}\right)\frac{\Gamma^{(3)}(r^2)}{\Gamma(r^2)} 
  + \frac13\left((u\cdot r)^2 + (u^\prime\cdot
  r)^2\right)\frac{\Gamma^{(4)}(r^2)}{\Gamma(r^2)}
  \right] \right] \nonumber\\
& & + \frac{u\cdot u^\prime\,\Gamma^{(1)}(r^2)+2u\cdot
   r\,u^\prime\cdot r\, \Gamma^{(2)}(r^2)}{2\Gamma^2(r^2)}
\left[- \left(\Gamma(u^2)+\Gamma(u^{\prime
    2})\right)\left(\left[S^{(2)}(r^2)\right]^2-1+C_F\Gamma(r^2)\right) 
\right. \nonumber\\
& & \left.
~~~~~~~~~~~~
+\left(\left[S^{(2)}(r^2)\right]^2\left(1+C_F\Gamma(r^2)\right)-1\right)
\left(\frac12\left(u-u^\prime\right)^2 \Gamma^{(1)}(r^2) + (u\cdot
r-u^\prime\cdot r)^2\Gamma^{(2)}(r^2)\right)
\right]~.  \label{eq:QG4thO}
\eea
This is the complete power-suppressed correction to ${\cal Q}^G$.  The
terms proportional to $r_i r_j r_m r_n$ which project onto $\sim \cos
4\phi$ are
\bea
{\cal G}^{ij , mn}  &=& r_i r_j r_m r_n\;  2 
\left( \frac{\Gamma^{(2)} (r^2) }{\Gamma(r^2)}  \right)^2 
\left[ 1 -\left[S^{(2)}(r^2)\right]^2 (1+C_F \Gamma(r^2)) \right] ~,\\
{\cal G}^{ij m ,n} &=& {\cal G}^{i,j mn} = 
	r_i r_j r_m r_n\;
	\frac13 \frac{\Gamma^{(4)} (r^2) }{\Gamma(r^2)} 
	\left(1-\left[S^{(2)}(r^2)\right]^2\right)
	- \frac12 {\cal G}^{ij ,mn}
\eea
and so
\begin{eqnarray}
{\cal G}^{ijmn}(r) &=& {\cal G}^{i,jmn} + {\cal G}^{ijm,n} - 
\frac{2}{3}{\cal G}^{ij, mn} \\
	 &=& r_i r_j r_m r_n \; \frac23  \left[ \frac{\Gamma^{(4)} (r^2)
  }{\Gamma(r^2)} \left(1-\left[S^{(2)}(r^2)\right]^2\right) - 5  
  \left( \frac{\Gamma^{(2)} (r^2) }{\Gamma(r^2)}  \right)^2 
	\left[ 1 -\left[S^{(2)}(r^2)\right]^2 (1+C_F \Gamma(r^2)) \right] 
	 \right]~.
\end{eqnarray}
Performing a Fourier transform like in Eq.~(\ref{eq:FTxGij}) and
projecting with $\mathfrak{P}_3$ we extract
\begin{eqnarray}
\Phi_2(x,q^2)  &=& - \frac{2 N_c}{\alpha_s} \mathfrak{P}_3^{ijmn} {\cal
  G}^{ijmn}(x,q^2)  \nonumber \\
&=& - \frac{N_c} {\sqrt{2} \, 3\pi \alpha_s} \frac{S_\perp}{(2\pi)^2} 
 \int d|r|\,  J_4 (|r|\, |q|)\, |r|^5 \nonumber\\
& &\times  \left[ \frac{\Gamma^{(4)} (r^2) }{\Gamma(r^2)} 
		 \left(1-\left[S^{(2)}(r^2)\right]^2\right) - 5  
  \left( \frac{\Gamma^{(2)} (r^2) }{\Gamma(r^2)}  \right)^2 
	\left[ 1 -\left[S^{(2)}(r^2)\right]^2 (1+C_F \Gamma(r^2)) \right] 
	 \right]~.
\end{eqnarray}
For the MV model, specifically,
\begin{eqnarray}
\Phi_2(q^2) &=& \frac{N_c}{\sqrt{2} \, 3\pi \alpha_s}   
\frac{S_\perp}{(2\pi)^2}   \int \frac{d|r|}{|r|^3} J_4(|r|\, |q|)
\left[ \frac{2}{ \ln \frac{1}{r^2\Lambda^2_{\rm IR}}} 
  \left\{1 - \exp\left( - \frac{Q_s^2 r^2}{4} \log
  \frac{1}{r^2\Lambda^2_{\rm IR}}   \right)  \right\} \right. \nonumber\\
&& +  \left. 
  \frac{5}{ \ln^2 \frac{1}{r^2 \Lambda^2_{\rm IR}}} \left\{  1 - 
  \exp\left( - \frac{Q_s^2 r^2}{4} \log  \frac{1}{r^2\Lambda^2_{\rm
      IR}}   \right)
  \left[ 1 + \frac{Q_s^2 r^2}{4} \log \frac{1}{r^2\Lambda^2_{\rm IR}}\right] 
	\right\}
\right]   \label{eq:Gauss_MV_Phi2}
\end{eqnarray}
For large $q\gg Q_s$ we have $\Phi_2(q^2) \to (N_c /\sqrt{2} \,
24\pi\alpha_s)\, (S_\perp/4\pi^2)\, Q_s^2$. For small $\Lambda_{\rm
  IR}\ll q\ll Q_s$ we have $\Phi_2(q^2) \sim (N_c/\alpha_s\log
Q_s^2/\Lambda^2_{\rm IR})\, S_\perp q^2$ with a coefficient that can
be determined numerically.

\begin{figure}
\includegraphics[width=0.45\linewidth]{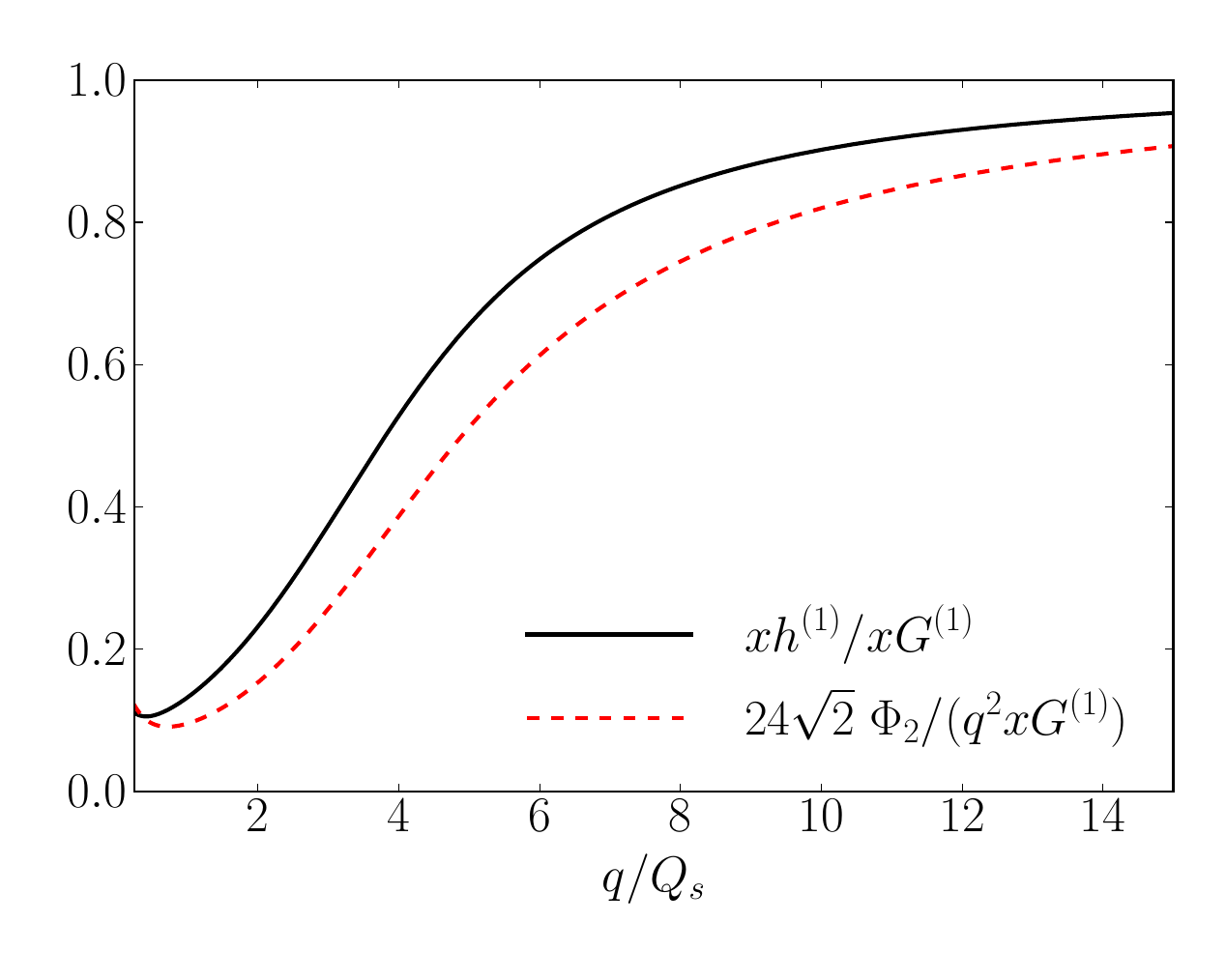}
\includegraphics[width=0.45\linewidth]{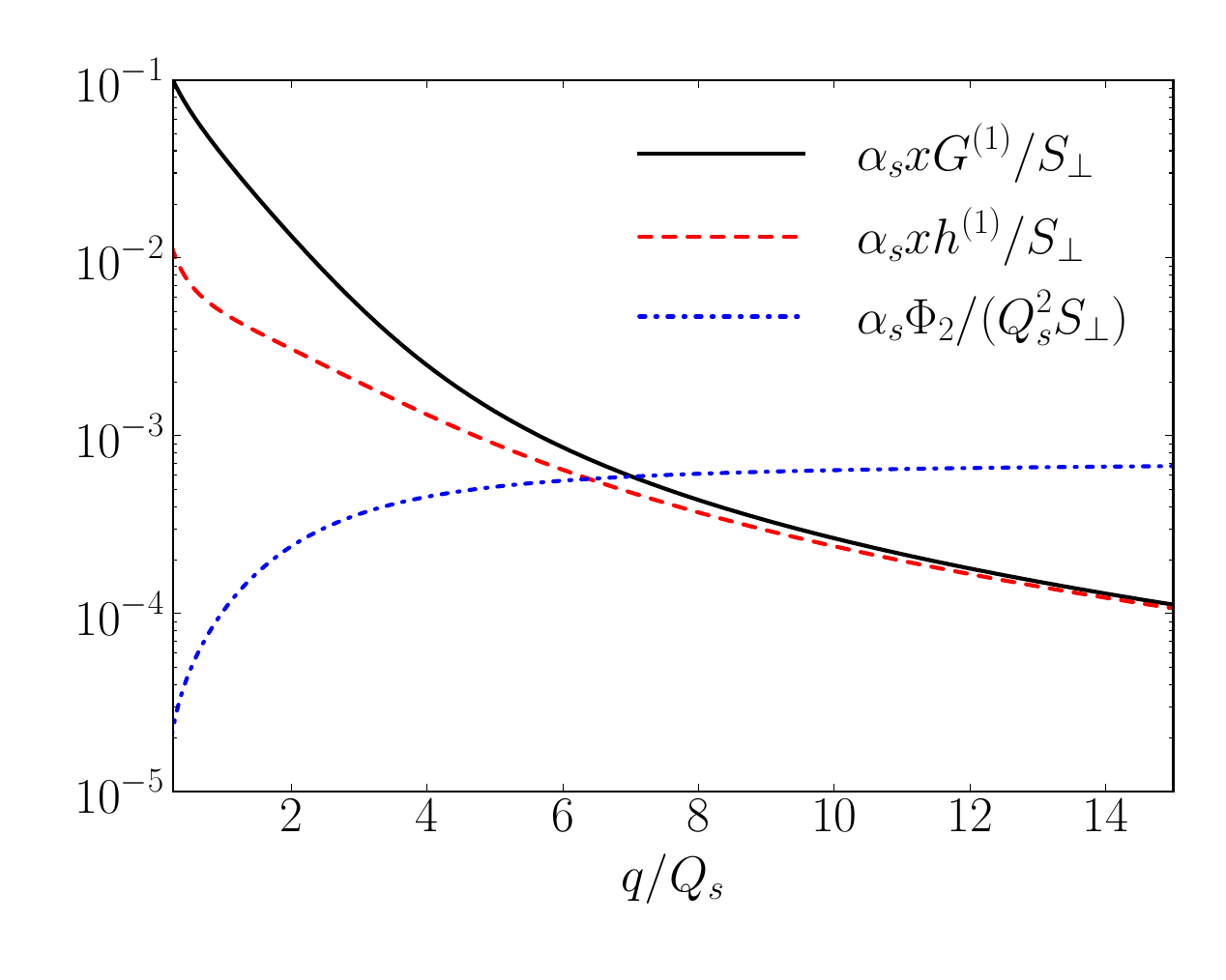}
\caption{The functions $xG^{(1)}(q^2)$, $xh^{(1)}(q^2)$ and
  $\Phi_2(q^2)$ in the MV model. These functions determine the
  amplitudes of the $\cos 2n\phi$ contributions to the dijet angular
  distributions for $n=0$, 1, 2, respectively. See text for details.}
\label{fig:MVfunctions}
\end{figure}
Figure~\ref{fig:MVfunctions} shows the functions $xG^{(1)}(q^2)$,
$xh^{(1)}(q^2)$ and $\Phi_2(q^2)$ in the MV model as written in
Eqs.~(\ref{eq:Gauss_MV_xh1},\ref{eq:Gauss_MV_xG1},\ref{eq:Gauss_MV_Phi2}). In
the numerical computations we have replaced $1/r^2 \Lambda^2_{\rm IR}
\to e + 1/r^2 \Lambda^2_{\rm IR}$ in the arguments of the logarithms
to ensure that they are $\ge1$ for all $r^2$. Also, we have used
$Q_s/\Lambda_{\rm IR} = 20$.

\section{Dijet cross section in DIS} \label{sec:DijetXsec}

At leading order the cross section for production of a $q\bar{q}$
dijet in DIS is given by~\cite{Dominguez:2011wm}
\begin{eqnarray}
\frac{d\sigma ^{\gamma_{T,L}^{\ast }A\rightarrow q\bar{q}X}}{d^2k_1dz_1d^2k_2dz_2}
&=&N_{c}\alpha _{\rm em}e_{q}^{2}\, \delta(p^+-k_1^+-k_2^+) \int
\frac{\text{d}^{2}u}{(2\pi)^{2}}\frac{\text{d}^{2}u^{\prime }}{(2\pi )^{2}}
\frac{\text{d}^{2}v}{(2\pi)^{2}}\frac{\text{d}^{2}v^{\prime }}{(2\pi )^{2}} 
  \notag \\
&&\times e^{-iP\cdot(u-u^{\prime })-iq\cdot (v-v^{\prime })}
{\cal Q}(u,u^{\prime },v,v^{\prime})
\sum_{\lambda\alpha\beta} \psi_{\alpha\beta}^{T, L \lambda}(u)\,
\psi_{\alpha\beta}^{T, L\lambda*}(u^{\prime }) 
~.\label{Eq:dis}
\end{eqnarray}
$k_1$ and $k_2$ denote the 2d transverse momenta of the quark and
anti-quark, respectively, and $P=(k_1-k_2)/2$, $q=k_1+k_2$. We assume
here that only the dijet is being detected while the azimuthal angle
of the electron is integrated over. If the azimuthal angle of the
electron can be measured then the dijet cross section could exhibit a
more involved angular dependence~\cite{Pisano:2013cya}.

In the ``correlation limit'' of roughly back to back jets $P^2\gg
q^2$. Using the $\gamma^*\to q\bar{q}$ splitting functions from the
literature, e.g.\ Ref.~\cite{Dominguez:2011wm}, and expanding ${\cal
  Q}$ to fourth order in $u$ or $u^\prime$ we obtain
\begin{eqnarray}
&& \frac{d\sigma ^{\gamma_{T}^{\ast }A\rightarrow q\bar{q}X}}{d^2k_1dz_1d^2k_2dz_2}
\nonumber\\
& & ~~~~~= 2 N_{c}\alpha _{em}e_{q}^{2}  \,(2\pi)^2 \delta(x_\gamma -z_1-z_2) 
\left(z_2^2 + z_1^2\right)
\int \frac{\text{d}^{2}u}{(2\pi)^{2}}\frac{\text{d}^{2}u^{\prime }}{(2\pi )^{2}}
\; e^{-i P\cdot(u-u^{\prime})} 
\nabla K_0(\epsilon_f u ) \cdot \nabla K_0(\epsilon_f
u^\prime)  \nonumber \\
&&~~~~~~~~~\times \left[
  u_{ i} u_{ j}^\prime {\cal G}^{i,j}(q) 
  + u_{ i} u_{ j}^\prime u_{ k}^\prime u_{ l}^\prime     {\cal
    G}^{i,jkl}(q)
  + u_{ i} u_{ j} u_{ k} u_{ l}^\prime     {\cal G}^{ijk,l}(q)
  + u_{ i} u_{ j} u_{ k}^\prime u_{ l}^\prime     
  {\cal G}^{ij, kl}(q) \right] \ ,\label{Eq:disT} \\
&&\frac{d\sigma ^{\gamma_{L}^{\ast }A\rightarrow q\bar{q}X}}{d^2k_1dz_1d^2k_2dz_2}
= 8N_{c}\alpha _{em}e_{q}^{2} \, (2\pi)^2 \delta(x_\gamma -z_1-z_2) 
z_1z_2\epsilon_f^2
\int \frac{\text{d}^{2}u}{(2\pi)^{2}}\frac{\text{d}^{2}u^{\prime }}{(2\pi )^{2}}
\; e^{-i P\cdot(u-u^{\prime})}
 K_0(\epsilon_f u )   K_0(\epsilon_f u^\prime) \notag \\
&&~~~~~~~~~\times \left[
   u_{ i} u_{ j}^\prime {\cal G}^{i,j}(q) 
   + u_{ i} u_{ j}^\prime u_k^\prime u_l^\prime {\cal G}^{i,jkl}(q) 
   + u_i u_j u_k u_l^\prime {\cal G}^{ijk,l}(q) 
   + u_i u_j u_k^\prime u_l^\prime {\cal G}^{ij,kl}(q) 
\right] ~.  \label{Eq:disL}
\end{eqnarray}
Here, $\epsilon_f^2 = z_1z_2 Q^2$ with $Q^2$ the virtuality of the
photon which is on the order of $P^2$.

The integrals over $u$ and $u^\prime$ can be performed using the
formulas collected in appendix~\ref{sec:Integrals}. The leading (in powers of $1/P^2$)
contributions proportional to $\cos 2n\phi$, for $n=0$, 1, 2, can be
summarized as
\begin{eqnarray}
&&\frac{d\sigma ^{\gamma_{T}^{\ast }A\rightarrow q\bar{q}X}}{d^2k_1dz_1d^2k_2dz_2}
\nonumber \\ 
&&~~~~~= \alpha_s  \alpha _{em}e_{q}^{2} \delta(x_\gamma -z_1-z_2) 
\left(z_1^2+ z_2^2\right)
\left[
\frac{P^4+\epsilon_f^4}{(P^2+\epsilon_f^2)^4}  
\left( xG^{(1)}(x,q^2)  - \frac{2\epsilon_f^2 P^2}{P^4+\epsilon_f^4}
xh^{(1)}(x,q^2) \cos 2\phi + {\cal O}\left(\frac1{P^2}\right)\right) 
\right. \nonumber\\
& & \left. \hspace{7cm}
-\frac{48 \epsilon_f^2 P^4}{\sqrt{2}\, (P^2+\epsilon_f^2)^6}  
\Phi_2(x,q^2)\cos 4 \phi \right]     \label{eq:Xsec_gammaT}\\
&&\frac{d\sigma ^{\gamma_{L}^{\ast }A\rightarrow q\bar{q}X}}{d^2k_1dz_1d^2k_2dz_2}
\nonumber \\ 
&& ~~~~~= 
8 \alpha_s \alpha _{em}e_{q}^{2} \delta(x_\gamma -z_1-z_2)
z_1z_2\epsilon_f^2 
\left[ \frac{ P^2}{(P^2+\epsilon_f^2)^4} 
\left( xG^{(1)}(x,q^2) + xh^{(1)}(x,q^2)\cos2\phi + 
     {\cal O}\left(\frac1{P^2}\right)
\right)\right. \nonumber\\
& & \left. \hspace{7cm}
+ \frac{48  P^4}{\sqrt{2}\, (P^2+\epsilon_f^2)^6} \Phi_2(x,q^2)  \cos 4 \phi
\right] ~.     \label{eq:Xsec_gammaL}
\end{eqnarray}
Here, $\cos\phi=\hat{q}\cdot \hat{P}$.  Note that the contribution
$\sim \cos 4 \phi$ is suppressed by $1/P^2$ relative to the isotropic
and $\sim \cos 2\phi$ pieces which are due to the $xG^{(1)}(x,q^2)$ and
$xh^{(1)}(x,q^2)$ TMDs.

\begin{figure}
\includegraphics[width=0.5\linewidth]{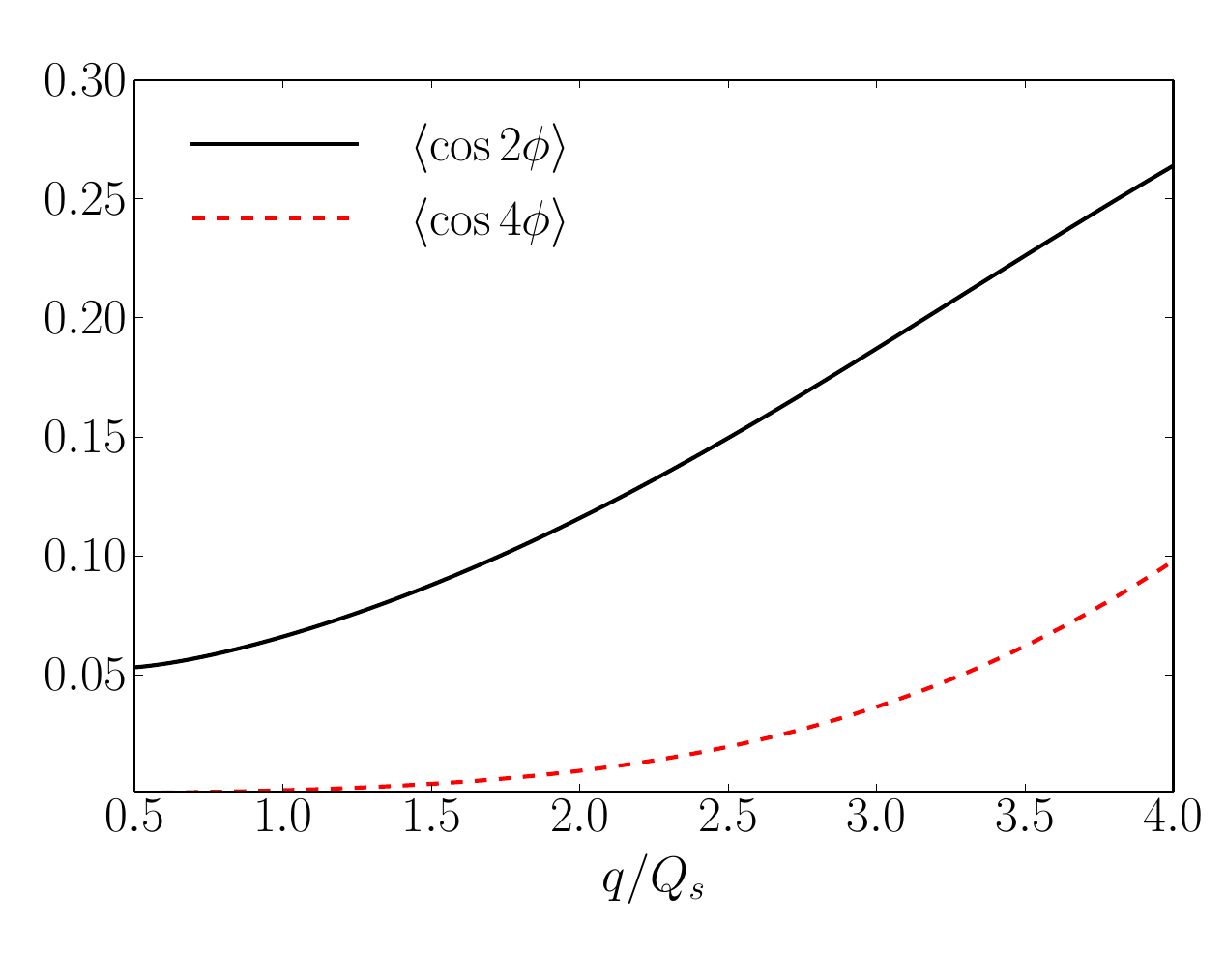}
\caption{$\left< \cos 2\phi\right>$ and $\left< \cos 4\phi\right>$ in
  $\gamma_L^*+A\to q+\bar{q}$ dijet production from the MV model.
  See text for details.}
\label{fig:cosnphi}
\end{figure}
Finally, we evaluate numerically the following angular averages for a
longitudinally polarized photon:
\be
\left< \cos 2\phi\right> = \frac12 \frac{xh^{(1)}(q^2)}{xG^{(1)}(q^2)}~~~,~~~
\left< \cos 4\phi\right> = \frac{24}{\sqrt{2}\, P^2}
\left(\frac{P^2}{P^2+\epsilon_f^2}\right)^2
\frac{\Phi_2(q^2)}{xG^{(1)}(q^2)}~.
\ee
We employ the MV model expressions for $xG^{(1)}(q^2)$,
$xh^{(1)}(q^2)$, and $\Phi_2(q^2)$ derived in the previous
section. The results are shown in Fig.~\ref{fig:cosnphi} assuming
$\surd P^2 = 4.5\, Q_s$, $z=0.5$ and $Q^2=P^2$. They confirm that for $q^2\ll
P^2$ the average $\cos 4\phi$ is substantially less than the average
$\cos 2\phi$ although it may be measurable at a future high-energy
electron-ion collider. For more quantitative estimates it is required,
however, to account for small-$x$ QCD evolution of these
functions. This has been done in Ref.~\cite{Dumitru:2015gaa} for
$xG^{(1)}(x,q^2)$ and $xh^{(1)}(x,q^2)$ and needs to be extended to
$\Phi_2(x,q^2)$.

\section{Summary} \label{sec:Summary}

In this paper we have considered the expansion of the quadrupole
operator
\be
S^{(4)}(x_1,x_2;x_2^{\prime },x_1^{\prime})
\equiv \frac{1}{N_c} \left< {\rm Tr}\,
V^\dagger(x_2)  V(x_1) V^\dagger(x_1^{\prime})  V(x_2^{\prime})
\right>
\ee
about the coincidence limits $u\equiv x_1-x_2\to0$, $u^\prime \equiv
x_1^\prime -x_2^\prime\to0$. At quadratic order it
becomes a two-point correlator of light-cone gauge fields~\cite{DQXY},
\be
u_i u^\prime_j\,
\left< {\rm Tr} \,
\left[  V^\dagger({v}) \, \frac{i}{g}\partial^i V({v})\right]\;
\left[  V^\dagger({v}^\prime) \, \frac{i}{g} \partial^j V({v}^\prime) \right]
\right>
\ee
which defines the Weizs\"acker-Williams and linearly polarized gluon
distributions. We have extended the expansion to fourth order in
$u/u^\prime$ which leads to more involved correlators of Wilson lines
and their derivatives,
c.f.\ Eqs.~(\ref{Eq:G22},\ref{Eq:G31a},\ref{Eq:G31b}). Furthermore, we
have obtained explicit analytic expressions in a Gaussian, large
$N_c$ approximation for the specific correlation function denoted as
$\Phi_2(x,q^2)$. This function gives rise to a $\sim \cos 4\phi$
azimuthal harmonic in dijet production. First qualitative estimates
obtained within a specific Gaussian model (McLerran-Venugopalan
model~\cite{McLerran:1993ni}) indicate that $\langle\cos 4\phi\rangle$
is much smaller than $\langle\cos 2\phi\rangle$ generated by the
distribution of linearly polarized gluons $xh^{(1)}(x,q^2)$, at least
in the nearly back to back ``correlation limit'' $P^2\gg q^2$.

\vspace*{1cm}
\begin{acknowledgments}
We appreciate insightful comments by E.~Aschenauer, A.~Kovner,
M.~Lublinsky and, especially, T.~Ullrich.  V.S.\ also thanks J.~Huang
and D.~Morrison for discussions and the organizers of the Spring 2016
fsPHENIX workshop where work on this paper was initiated.

A.D.\ gratefully acknowledges support by the DOE
Office of Nuclear Physics through Grant No.\ DE-FG02-09ER41620; and
from The City University of New York through the PSC-CUNY Research
grant 69362-00~47.

\end{acknowledgments}

\appendix
\section{Useful Integrals} \label{sec:Integrals}
We start from the well known integral 
\begin{equation}
	\int \frac{d^2 u }{(2\pi)^2} \exp(-i u\cdot P)
 K_0(\epsilon_f u) 	= \frac{1}{2\pi} \frac{1}{P^2 + \epsilon_f^2}~.   
	\label{Eq:K0Int}
\end{equation}
Taking a derivative with respect to $P_{ i}$ gives
\begin{equation}
  \int \frac{d^2 u }{(2\pi)^2}  \exp(-i u\cdot P) 
  u_{ i}   K_0(\epsilon_f u)	= \frac{1}{2\pi i} \frac{2 P_{
      i}}{(P^2 + \epsilon_f^2)^2}~.
	\label{Eq:K0uInt}
\end{equation}
Repeating this procedure one finds
\begin{equation}
  \int \frac{d^2 u }{(2\pi)^2}  \exp(-i u\cdot P) 
  u_{ i} u_{ j}  K_0(\epsilon_f u) =
  \frac{1}{2\pi} \frac{2}{(P^2+\epsilon_f^2)^2} \left[ \delta_{ij} -
    \frac{4P_{ i}P_{ j}}{P_{}^2+\epsilon_f^2} \right]
  \label{Eq:K0uuint}
\end{equation}
and 
\begin{equation}
  \int \frac{d^2 u }{(2\pi)^2}  \exp(-i u\cdot P) 
  u_{ i} u_{ j} u_{ k}   K_0(\epsilon_f u) =
  \frac{1}{2\pi i}  \frac{8}{(P^2+\epsilon_f^2)^3} 
  \left[ P_{ i}{\delta}_{jk}  + P_{ j} {\delta}_{ik} +P_{ k} {\delta}_{ij} 
- \frac{6P_iP_jP_k}{P^2+\epsilon_f^2} \right]~.
	\label{Eq:K0uuuint}
\end{equation}

For transverse photon polarization we need
\bea
  \int \frac{d^2 u }{(2\pi)^2}  \exp(-i u\cdot P) 
  \frac{\partial}{\partial u_l} K_0(\epsilon_f u) &=& - 
  \int \frac{d^2 u }{(2\pi)^2}   \left(\frac{\partial}{\partial u_l}
  \exp(-i u\cdot P)
  \right) K_0(\epsilon_f u) = - \frac{1}{2\pi i } \frac{P_{
      l}}{P^2+\epsilon_f^2} ~,
	\label{Eq:gradK0int}\\
  \int \frac{d^2 u }{(2\pi)^2}  \exp(-i u\cdot P) 
  u_{ i} \frac{\partial}{\partial u_l} K_0(\epsilon_f u)	&=& 
  -    \frac{1}{2\pi }   \frac{1}{P^2+\epsilon_f^2}
  \left({\delta}_{il}-\frac{2P_iP_l}{P^2+\epsilon_f^2}\right)~,
  \label{Eq:gradK0uint} \\
  \int \frac{d^2 u }{(2\pi)^2}  \exp(-i u\cdot P) 
  u_{ i} u_{ j} \frac{\partial}{\partial u_l} K_0(\epsilon_f u) &=& 
  -   \frac{1}{2\pi i }   \frac{2}{(P^2+\epsilon_f^2)^2}
  \left(
 \delta_{ij} P_{ l}  + 
 {\delta}_{il} P_{ j} +{\delta}_{jl} P_{ i}
- \frac{4P_lP_iP_j}{P^2+\epsilon_f^2}
\right)~,
\label{Eq:gradK0uuint} \\
  \int \frac{d^2 u }{(2\pi)^2}  \exp(-i u\cdot P) 
  u_{ i} u_{ j} u_{ k} \frac{\partial}{\partial u_l}
  K_0(\epsilon_f u) &=& 
  -   \frac{1}{2\pi }   \frac{2}{(P_{}^2+\epsilon_f^2)^2} \nonumber\\ 
&& \hspace{-7cm} \times \left( 
  \delta_{il}\delta_{jk} + 
  \delta_{ij}\delta_{kl} + 
  \delta_{ik}\delta_{jl} 
  - 4 \frac{  
    P_{ j}   P_{ k } \delta_{il} + 
    P_{ l}   P_{ k } \delta_{ij} + 
    P_{ i}   P_{ k } \delta_{jl} + 
    P_{ j}   P_{ l } \delta_{ik} + 
    P_{ i}   P_{ j } \delta_{lk} + 
    P_{ i}   P_{ l } \delta_{jk}  
  }{ P_{}^2+\epsilon_f^2}
  + 24 \frac{ P_{ i}   P_{ j }  P_{ l}   P_{ k } }{(
    P_{}^2+\epsilon_f^2)^2  }
  \right)
  \label{Eq:gradK0uuintb}
\eea

\section{Power corrections to the isotropic and $\cos 2\phi$
  contributions} \label{sec:PowerCorrec_xG_xh}

In this appendix we derive the leading power corrections to the
isotropic and $\sim \cos 2\phi$ contributions in
Eqs.~(\ref{eq:Xsec_gammaT}, \ref{eq:Xsec_gammaL}). These are
suppressed by one power of $P^2$ but logarithmically enhanced by a
factor of $\log P^2/\Lambda^2$, and arise from the terms in the
expansion of the quadrupole ${\cal Q}^G$ in (\ref{eq:QG4thO}) which
involve $\Gamma(u^2)$ or $\Gamma(u'^2)$.

First we write
\be
\Gamma(u^2) = u^2 \hat \Gamma(u^2)~~~~,~~~~
\hat\Gamma(u^2)=\frac{Q_s^2}{4C_F}\log \frac{1}{u^2\Lambda^2}~.
\ee
The prefactor of the logarithm in $\hat\Gamma(u^2)$ is specific to the
MV model but $\hat\Gamma(u^2)\sim \log {1}/{u^2\Lambda^2}$ applies at
small $u$ even when the gluon distribution acquires an anomalous
dimension.

Now we extend the expansion about small $u$, $u'$ from
Eqs.~(\ref{Eq:Qexpansion}, \ref{eq:Xsec_gammaT}, \ref{eq:Xsec_gammaL})
as follows:
\bea
& &
 u_{ i} u_{ j}^\prime {\cal G}^{i,j}(q)  \nonumber\\
  && 
  + u_{ i} u_{ j}^\prime u_{ k}^\prime u_{ l}^\prime
       \log\frac{1}{u^{\prime 2}\Lambda^2}
      {\cal G}^{i,jkl}_\text{log}(q)
 + u_{ i} u_{ j} u_{ k} u_{ l}^\prime
       \log\frac{1}{u^2\Lambda^2}
      {\cal G}^{ijk,l}_\text{log}(q)
      + u_{ i} u_{ j} u_{ k}^\prime u_{ l}^\prime
      \log\frac{1}{u^{\prime 2}\Lambda^2}\log\frac{1}{u^2\Lambda^2}
  {\cal G}^{ij, kl}_\text{log}(q)  ~.
\eea
Here, we have only listed the leading power and the log-enhanced power corrections.
From Eq.~(\ref{eq:QG4thO}) one can read off
\bea
    {\cal G}^{i,jkl}_\text{log}(r) &=& \frac{Q_s^2}{4C_F} \left(1
    -\left[S^{(2)}(r^2)\right]^2 -C_F \Gamma(r^2) \right)\delta^{kl}
         \frac{\delta^{ij}\Gamma^{(1)} (r^2)+2r^ir^j\Gamma^{(2)}
           (r^2)}{2\Gamma^2(r^2)} ~,\\
    {\cal G}^{ijk,l}_\text{log}(r) &=&  \frac{Q_s^2}{4C_F} \left(1
    -\left[S^{(2)}(r^2)\right]^2 -C_F \Gamma(r^2) \right)\delta^{jk}
         \frac{\delta^{il}\Gamma^{(1)} (r^2)+2r^ir^l\Gamma^{(2)}
           (r^2)}{2\Gamma^2(r^2)} ~,\\
    {\cal G}^{ij, kl}_\text{log}(r)  &=& \frac{1}{4}\delta^{ij}
    \delta^{kl} C_F^2 \left(\frac{Q_s^2}{4C_F}\right)^2 ~.
\eea
We now perform the integrals over $u$ and $u'$ written in
Eqs.~(\ref{eq:Xsec_gammaT}, \ref{eq:Xsec_gammaL}) using the formulas
given in appendix~\ref{sec:Integrals}. For the terms involving
logarithms those expressions have to be multiplied by $\log
1/u^2\Lambda^2$ with $1/u$ replaced by $(P^2+\epsilon_f^2)/\epsilon_f$
(in the leading $\log P/\Lambda$ approximation). The final step is to
perform a Fourier transform w.r.t.\ $r=v-v'$, which is conjugate to
$q$, and an integration over $(v+v')/2$, which gives a factor of
transverse area. We then find that in Eq.~(\ref{eq:Xsec_gammaT}) we
have to replace\footnote{We suppress a contribution proportional to
  $\delta^2(q)\log^2({P^2+\epsilon_f^2})/({\Lambda \epsilon_f})$
  which arises from the $r$-independent ${\cal G}^{ij, kl}_\text{log}$
  since here we do not address the gluon distributions integrated over
  $q$.}
\bea
xG^{(1)}(x,q^2) &\to& xG^{(1)}(x,q^2) -
 Q_s^2 \frac{\epsilon_f^2 \left( (P^2-\epsilon_f^2)^2+2P^4 \right)
}{\left(P^2+\epsilon_f^2\right)^2 (P^4+\epsilon_f^4)} \log
\left(\frac{P^2+\epsilon_f^2}{\Lambda  \epsilon_f}\right) x\tilde
G^{(1)}(x,q^2)~,     \label{eq:xG_replaceT}\\
xh^{(1)}(x,q^2) &\to& xh^{(1)}(x,q^2) -
Q_s^2 \frac{ 2\epsilon_f^2 - P^2}
       {\left(P^2+\epsilon_f^2\right)^2}
       \log \left(\frac{P^2+\epsilon_f^2}{\Lambda  \epsilon_f}\right)
       x\tilde h^{(1)}(x,q^2)~.     \label{eq:xh_replaceT}
\eea
Similarly, in Eq.~(\ref{eq:Xsec_gammaL}),
\bea
xG^{(1)}(x,q^2) &\to& xG^{(1)}(x,q^2) - 2Q_s^2 \frac{2 \epsilon_f^2 -P^2
}{\left(P^2+\epsilon_f^2\right)^2} \log
\left(\frac{P^2+\epsilon_f^2}{\Lambda  \epsilon_f}\right) x\tilde
G^{(1)}(x,q^2)~,    \label{eq:xG_replaceL}\\
  xh^{(1)}(x,q^2) &\to& xh^{(1)}(x,q^2) -  2Q_s^2 \frac{2 \epsilon_f^2 -P^2}
       {\left(P^2+\epsilon_f^2\right)^2}
       \log \left(\frac{P^2+\epsilon_f^2}{\Lambda  \epsilon_f}\right)
       x\tilde h^{(1)}(x,q^2)~, \label{eq:xh_replaceL}
\eea
with
\bea
x\tilde G^{(1)}(x,q^2) &=& \frac{8 N_c}{\alpha_s} \frac{S_\perp}{(2\pi)^2}  
\int \frac{d^2r}{(2\pi)^2}  e^{-iq \cdot r}
\left(C_F \Gamma(r^2)+\left[S^{(2)}(r^2)\right]^2 -1 \right)
\frac{ \Gamma^{(1)}(r^2) + r^2 \Gamma^{(2)}(r^2)}{C_F \Gamma^2 (r^2)} \\
  x\tilde h^{(1)}(x,q^2) &=&  \frac{8 N_c}{\alpha_s}  \frac{S_\perp}{(2\pi)^2}  
       \int \frac{d^2r}{(2\pi)^2}  e^{-iq \cdot r}  \cos(2\phi_r)
       \left(C_F \Gamma(r^2)+\left[S^{(2)}(r^2)\right]^2 -1 \right) 
	\frac{r^2 \Gamma^{(2)}(r^2)}{C_F \Gamma^2 (r^2)} ~.
\eea
At high transverse momentum, $q^2\gg Q_s^2$, the functions $x\tilde
G^{(1)}(x,q^2)$ and $x\tilde h^{(1)}(x,q^2)$ approach
$xG^{(1)}(x,q^2)$ and $xh^{(1)}(x,q^2)$, respectively. In that limit,
the corrections in Eqs.~(\ref{eq:xG_replaceT}, \ref{eq:xh_replaceT},
\ref{eq:xG_replaceL}, \ref{eq:xh_replaceL}) are of order
$(Q_s^2/P^2)\, \log P^2/\Lambda^2$. Assuming, for example, $P^2/10=
\epsilon_f ^2/10 = 20\Lambda^2=q^2=Q_s^2$ leads to corrections of about
8\% for transverse polarization, and twice that for a longitudinal photon.

For comparison, the power correction that generates the $\sim\cos
4\phi$ contribution in Eqs.~(\ref{eq:Xsec_gammaT},
\ref{eq:Xsec_gammaL}) is of order $q^2/P^2$. To see this, write
$\tilde \Phi_2(x,q^2) = \Phi_2(x,q^2) /q^2$; this function has the
same dimension and the same $\sim 1/q^2$ fall off at high transverse
momentum as $xG^{(1)}(x,q^2)$ and $xh^{(1)}(x,q^2)$. Its
prefactor in Eqs.~(\ref{eq:Xsec_gammaT}, \ref{eq:Xsec_gammaL}) is then
suppressed by one power of $q^2/P^2$ as compared to the prefactors of
$xG^{(1)}(x,q^2)$ and $xh^{(1)}(x,q^2)$.

\end{document}